\documentclass{article}
\usepackage{graphicx}
\begin{document}
%+Title
\title{Quantum cosmology with varying speed of light: canonical approach}
\author{P. Pedram\thanks{pouria.pedram@gmail.com}, S. Jalalzadeh\thanks{Email: s-jalalzadeh@sbu.ac.ir}
\\ {\small Department of Physics, Shahid Beheshti University,
Evin, Tehran 19839, Iran}}
\date{\today}
\maketitle \baselineskip 24pt
%-Title

%+Abstract
\begin{abstract}
We investigate $(n+1)$--dimensional cosmology with varying speed of
light. After solving corresponding Wheeler-DeWitt equation, we
obtain exact solutions in both classical and quantum levels for
($c\,$--$\Lambda$)--dominated Universe. We then construct the
``canonical'' wave packets which exhibit a good classical and
quantum correspondence. We show that arbitrary but appropriate
initial conditions lead to the same classical description. We also
study the situation from de-Broglie Bohm interpretation of quantum
mechanics and show that the corresponding Bohmian trajectories are
in good agreement with the classical counterparts.
\end{abstract}
%-Abstract
\section{Introduction}\label{sec1}
In recent years the varying speed of light theories (VSL) have
attracted much attentions
\cite{mof1,time,b,bm1,bm2,bm3,bm4,mof2,cly1,cly2,cly3,drum,av1,av2,harko,kir,alex,bass}
(for a comprehensive review see \cite{magu}). These theories
proposed by Moffat \cite{mof1} and Albrecht and Magueijo
\cite{time}, in which light is traveling faster in the early periods
of the existence of the Universe, could be considered as an
alternative to the inflation scenario. It has been shown that the
horizon, flatness, and cosmological constant problems can be solved
in these models. Moreover, homogeneity and isotropy problems may
find their appropriate solutions through this mechanism \cite{time}.
Recently, an interesting discussion on the foundations of VSL
theories and the conceptual problems arising from the meaning of
varying speed of light have been done by Ellis, Magueijo and Moffat
\cite{note,comment}.

It is shown that it is possible to generalize these ideas to
preserve the general covariance and local lorentz invariance
\cite{mague}. They have the merit of retaining only those aspects of
the usual definitions that are invariant under unit transformations
and which can therefore represent the outcome of an experiment. This
can be done by introducing a time-like coordinate $x^0$ which is not
necessarily equal to $ct$. In terms of $x^0$ and $\vec{x}$, we have
local Lorentz invariance and general covariance. The physical time
$t$, can only be defined when $dx^0/c$ is integrable.

Some authors have studied quantum cosmological aspects of VSL models
\cite{har,yu,shojai}. In particular, Shojai \textit{et al}
\cite{shojai} consider FRW quantum cosmological models with varying
speed of light in the presence of cosmological constant. They solved
the corresponding Wheeler-DeWitt (WDW) equations exactly and found
the eigenfunctions. Then, they used these eigenfunctions to
construct the Bohmian trajectories via de-Broglie Bohm
interpretation of quantum mechanics. As they have truly stated, the
Bohmian trajectories highly depend on the wave function of the
system and various linear combinations of eigenfunctions lead to
different Bohmian trajectories.

On the other hand, a legitimate question which arises is, how we can
construct a specific wave packet which completely corresponds to its
unique classical counterpart? First let us explain what we expect
from classical-quantum correspondence. A good classical-quantum
correspondence means that the wave packet centered around the
classical path, the crest of the wave packet should follow as
closely as possible the classical path, and to each distinct
classical path there should correspond a wave packet with the above
properties. The first part of this condition implies that the
initial wave function should consist of a few localized pieces.
Secondly, one expects the square of the wave packet describing a
physical system to posses a certain degree of smoothness.

Here, we use the method that is presented in Ref.~\cite{pedram} to
construct the wave packets with the above properties which so called
``canonical'' wave packets. Furthermore, we use de-Broglie Bohm
interpretation to find its corresponding Bohmian trajectories and
compare them with the classical ones. We will show that the
resulting Bohmian trajectories which are obtained from canonical
wave packets, are in good agreement with the classical counterparts.
It is worth to mention that since time is absent in quantum
cosmology, some other methods like Schutz's formalism also can be used to
recover the notion time \cite{Schutz1,Schutz2}.

The paper is organized as follows: in Sec.~\ref{sec2}, we present
the action in $n+1$ dimensions and reduce it to a more simpler form
using appropriate transformations. In Sec.~\ref{sec3}, we quantize
the model and obtain the exact solutions of WDW equation. Then we
construct the canonical wave packets using the prescription stated
in Ref. \cite{pedram}. In Sec.~\ref{sec4}, we find the corresponding
Bohmian trajectories and compare the classical and quantum
mechanical solutions. In Sec.~\ref{sec5}, we state our conclusions.

\section{The model}\label{sec2}
Let us start from the Einstein-Hilbert action for varying speed of
light theory \cite{magu,mague,shojai} generalized in $n+1$
dimensions
\begin{equation}\label{action}
S=\int d^{n+1}x \sqrt{-g}\left( e^{\alpha\psi}\left ( {\cal
R}-2\Lambda(\psi)-\kappa\nabla_\mu\psi\nabla^\mu\psi\right )
+e^{\beta\psi}{\cal L}_m(\phi_i,\partial_\mu\phi_i)\right )
\end{equation}
where $\psi=\log(c/c_0)$ is the scalar field and $c_0$ is a constant
velocity. Units are chosen such that the factor $16\pi G/c_0^4$
becomes equal to one. We have also consider a dynamical term for the
velocity of light with a dimensionless coupling constant $\kappa$,
and $\phi_i$ represent matter fields. Note that for $n=3$,
$\alpha=4$, and $\beta=0$ this theory is nothing but a unit
transformation applied to Brans-Dicke theory \cite{magu}. Particle
production and second quantization for this model have been
discussed in \cite{mague} and Black hole solutions are also studied
\cite{mague2}. Fock--Lorentz space--time \cite{80,81} as the ``free''
solution, and fast--tracks as solutions driven by cosmic strings
\cite{mague} are other interesting issues which have been
investigated. In this formalism, we use an ``$x^0$'' coordinate,
with dimension of length rather than time. With this choice, $c$
appears nowhere in the usual definitions of differential geometry,
which may therefore still be used. In fact, $x^0$ is not equal to
$ct$ and since $c$ is a field, $c\,dt$ is not necessarily
integrable. Therefore, definition of physical time is only possible
when $dx^0/c$ is integrable \cite{mague}.

Let us consider a $(n+1)$ dimensional FRW Universe, since we want to
deal with the cosmological problem. In this situation, the
Lagrangian (\ref{action}) becomes
\begin{eqnarray} \label{5}
{\cal
L}=a^{n}e^{\alpha\psi}\left[-n(n-1)\left(\frac{\dot{a}}{a}\right)^{2}-
2\alpha
n\dot{\psi}\left(\frac{\dot{a}}{a}\right)+n(n-1)\frac{k}{a^{2}}-\kappa\dot{\psi}^{2}
-2\Lambda(\psi)\right]+a^{n}e^{\beta\psi}{\cal
L}_{m}(\phi_i,\partial_\mu\phi_i)\,.
\end{eqnarray}
where $a$ is the scale factor and the constant $k$ is the spatial
curvature constant which can be $k=+1,-1,0$ for spatially closed,
open and flat cosmological models, respectively. Since recent
observations are in agreement with the assumption of flat Universe,
we assume $k=0$.

To simplify the Lagrangian we can use the change of variable
$b=e^{-\varphi}$ which leads to
\begin{eqnarray}
\psi=\ln(b),  \hspace{3cm}\, \dot{\psi}=\frac{\dot{b}}{b}
\end{eqnarray}
In terms of $a$ and $b$, the Lagrangian for a
($c\,$--$\Lambda$)-dominated Universe (${\cal L}_{m}=0$) can be
written as
\begin{eqnarray}
{\cal L}=-b^\alpha\left[n(n-1)\dot{a}^{2}a^{n-2}- 2\alpha
n\frac{\dot{b}}{b}\dot{a}a^{n-1}+\kappa\left(\frac{\dot{b}}{b}\right)^2a^n
+2a^{n}\Lambda(b)\right].\label{lag}
\end{eqnarray}
Now, we define new variables
\begin{eqnarray}\label{uvar}
u+v&=&A a^{\alpha'}b^{\beta'},\\\label{vvar}
u-v&=&a^{\gamma'}b^{\eta'},
\end{eqnarray}
%and their derivatives
%\begin{eqnarray}
%\dot{u}+\dot{v}&=&A(\alpha' a^{\alpha'-1}b^{\beta'}\dot{a}+\beta' a^{\alpha'}b^{\beta'-1}\dot{b})\\
%\dot{u}-\dot{v}&=&\gamma' a^{\gamma'-1}b^{\eta'}\dot{a}+\eta'
%a^{\gamma'} b^{\eta'-1}\dot{b}
%\end{eqnarray}
where $\alpha'$, $\beta'$, $\gamma'$, $\eta'$ and $A$ are constants.
Since we are interested to decouple the variables in the Lagrangian,
we choose the constants to reduce the kinetic part of the Lagrangian
to $\dot{u}^2-\dot{v}^2$. This means
\begin{eqnarray}\nonumber
\dot{u}^2-\dot{v}^2&=&A\left[\alpha'\gamma'
a^{\alpha'+\gamma'-2}b^{\eta'+\beta'}\dot{a}^2+(\alpha' \eta'+\beta'
\gamma')a^{\alpha'+\gamma'-1}b^{\eta'+\beta'-1}\dot{a}\dot{b}+\beta'
\eta'
a^{\gamma'+\alpha'}b^{\eta'+\beta'-2}\dot{b}^2\right]\\
&=&b^{\alpha}\left[-n(n-1)\dot{a}^{2}a^{n-2}- 2\alpha
n\frac{\dot{b}}{b}\dot{a}a^{n-1}-\kappa\left(\frac{\dot{b}}{b}\right)^2a^n
\right].
\end{eqnarray}
Which leads to the following equations
\begin{equation}
\left\{
\begin{array}{llllll}
\alpha'+\gamma' &=&n ,\\
  \\
\beta'+\eta'&=&\alpha,  \\
 \\
A\alpha'\gamma' &=&-n(n-1),\\
  \\
A\eta'\beta' &=&-\kappa,\\
 \\
A(\alpha'\eta'+\beta'\gamma') &=&-2\alpha n .\\
 \\
\end{array}
\right.
\end{equation}
%\begin{eqnarray}
%\alpha'+\gamma'&=&n\\
%\beta'+\eta'&=&\alpha\\
%A\alpha'\gamma'&=&-n(n-1)\\
%A\eta'\beta'&=&-\kappa\\
%A(\alpha'\eta'+\beta'\gamma')&=&-2\alpha n
%\end{eqnarray}
Finally, in terms of $u$ and $v$ the Lagrangian (\ref{lag}) takes
the form
\begin{eqnarray}\label{new-L}
{\cal L}=\dot{u}^2-\dot{v}^2-\frac{2}{A}(u^2-v^2)\Lambda(u,v).
\end{eqnarray}
The corresponding Hamiltonian can be easily obtained as
\begin{eqnarray}\label{new-H}
{\cal H}=\frac{p^2_u-p^2_v}{4}+\frac{2}{A}(u^2-v^2)\Lambda(u,v),
\end{eqnarray}
where $p_u=\frac{\displaystyle \partial{\cal
L}}{\partial\displaystyle \dot{u}}$ and $p_v=\frac{\displaystyle
\partial{\cal L}}{\displaystyle \partial\dot{v}}$. Therefore, the classical
equations of motion for $u$ and $v$ directions are
\begin{eqnarray}
\ddot{u}&=&\frac{\displaystyle1}{\displaystyle A}\left[2u+(u^2-v^2)\frac{\displaystyle\partial }{\displaystyle\partial u}\right]\Lambda(u,v),  \\
\ddot{v}&=&\frac{\displaystyle1}{\displaystyle
A}\left[2v+(v^2-u^2)\frac{\displaystyle\partial
}{\displaystyle\partial v}\right]\Lambda(u,v),\\
0&=&\dot{u}^2-\dot{v}^2+\frac{2}{A}(u^2-v^2)\Lambda(u,v),
\end{eqnarray}
where the last equation is zero energy condition. For $\psi$
independent cosmological constant ($\Lambda(\psi)=\Lambda$), these
equations represent a two dimensional Simple Harmonic Oscillator
(SHO) with the same frequency in each direction. In this case, the
classical trajectories are circles with arbitrary radius
(\textit{i.e.} $\xi$) in configuration space.% Moreover, as can be
%seen from equations (\ref{uvar},\ref{vvar}) singularities ($a=0$)
%happen in $u=v$ and/or $u=-v$ directions.

\section{Quantum cosmology and wave packets}\label{sec3}
Let us now turn to the study of quantum cosmology of the model
presented above. The Hamiltonian can then be obtained upon
quantization $p_u\rightarrow
-i\frac{\displaystyle\partial}{\displaystyle\partial u}$ etc., one
arrives at the WDW equation describing the corresponding quantum
cosmology
\begin{eqnarray}
{\cal H} \Psi(u,v)=\left\{-\frac{\partial^{2}}{\partial
u^{2}}+\frac{\partial^{2}}{\partial v^{2}}+{\omega}^2(u^{2}-v^{2})
\right\}\Psi(u,v)=0. \label{4.3}
\end{eqnarray}
Where, $\omega=\sqrt{\frac{{\displaystyle 8\Lambda}}{\displaystyle
A}}$. Note that, the appropriate transformations
(\ref{uvar},\ref{vvar}) prevent us from facing factor ordering
problem which usually arises \cite{shojai}. This equation is
separable in the minisuperspace variables and a solution can be
written as
\begin{equation}
\Phi_n(u,v)=\psi_{n}(u)\psi_{n}(v)
\label{4.4}
\end{equation}
where
\begin{eqnarray}
\psi_{n}(x)=\left(\frac{\omega}{\pi}\right)^{1/4}\left[\frac{H_{n}\left(
\sqrt{\omega}x\right)}{\sqrt{2^{n}n!}}\right]e^{ -\omega x^{2}/2}
\label{4.5}
\end{eqnarray}
In these expressions $ H_{n}(x)$ is a Hermite polynomial and the
orthonormality and completeness of the basis functions follow from
those of the Hermite polynomials.

\begin{figure}
\centerline{\begin{tabular}{ccc}
\includegraphics[width=8cm]{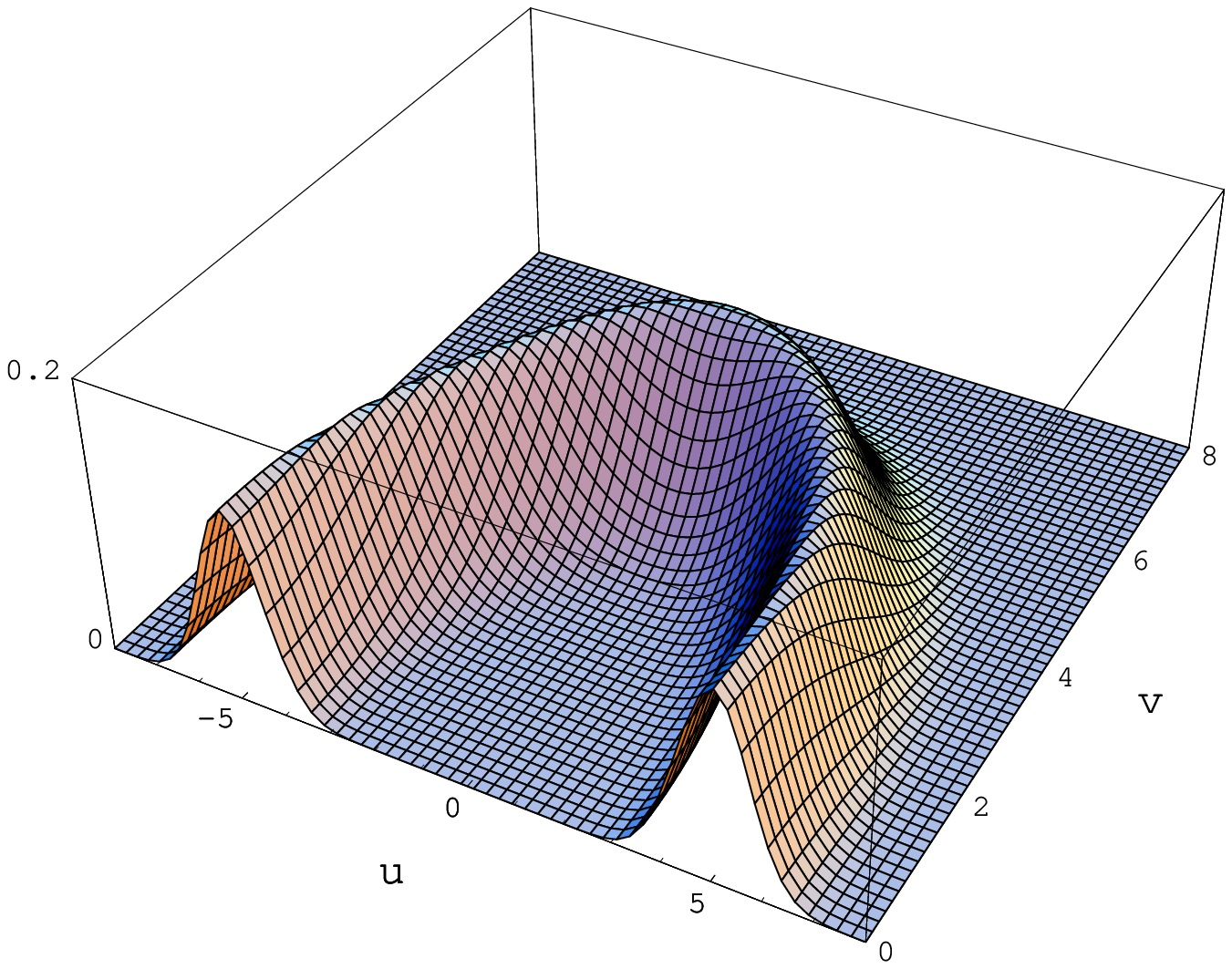}
 &\hspace{2.cm}&
\includegraphics[width=8cm]{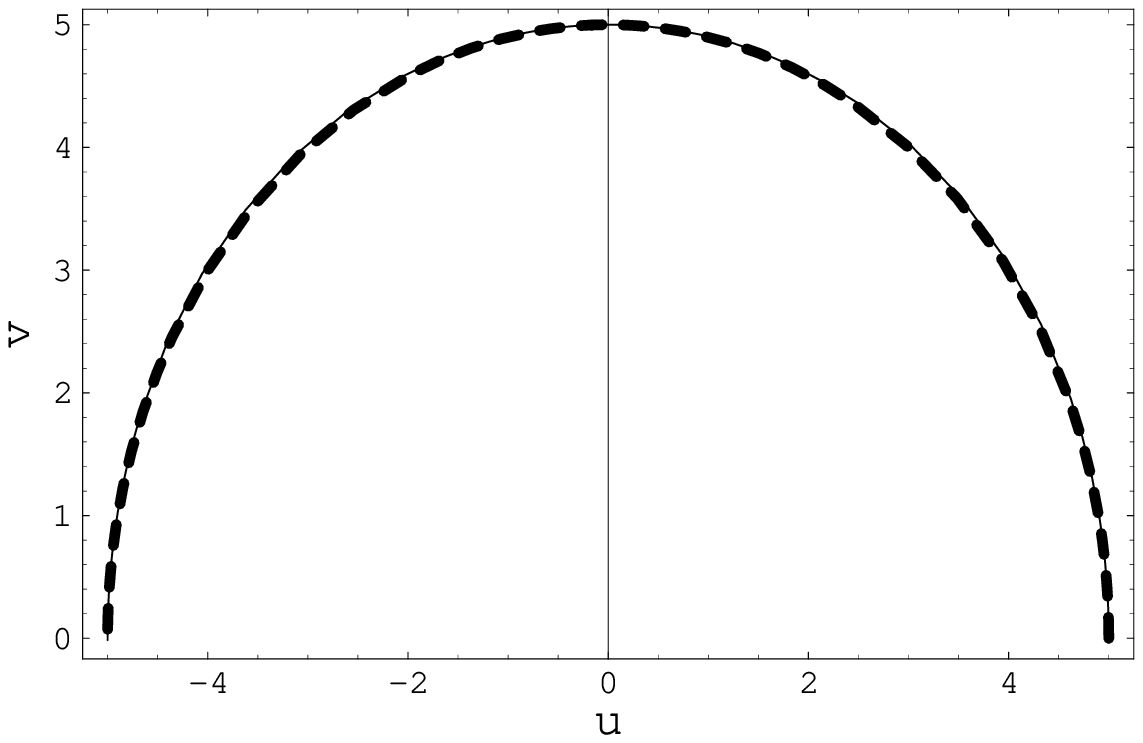}
\end{tabular}}
\caption{Left, the square of the wave packet $| \psi(u,v)|^2$ for
$A^*(n)=\frac{\,\chi^n}{\,{\sqrt{2^n\,n!}}}e^{-\chi^2/4}$ and
$\chi=5$. Right, the classical (dashed line) and bohmian (solid
line) trajectories.} \label{fig1}
\end{figure}

\begin{figure}
\centerline{\begin{tabular}{ccc}
\includegraphics[width=8cm]{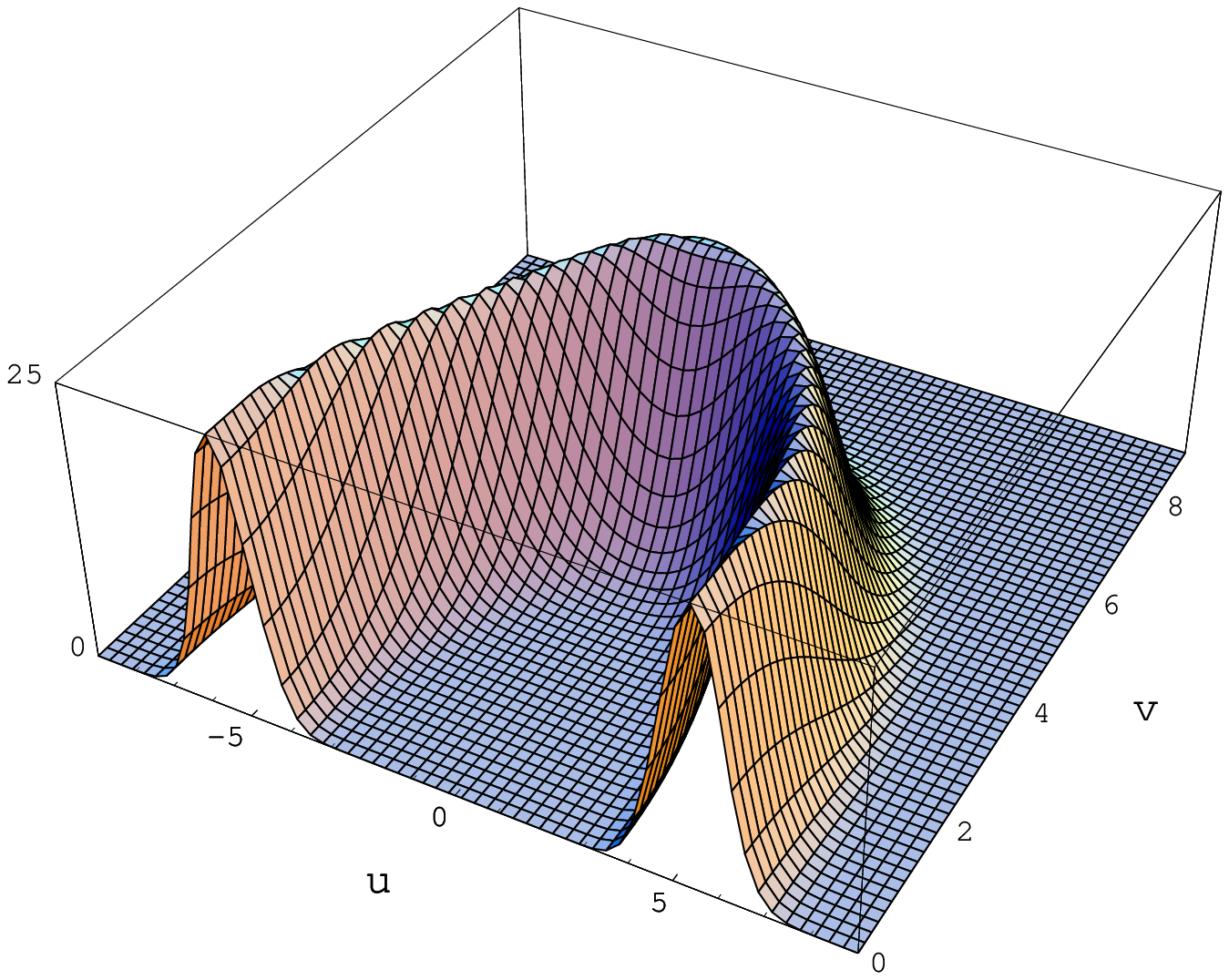}
 &\hspace{2.cm}&
\includegraphics[width=8cm]{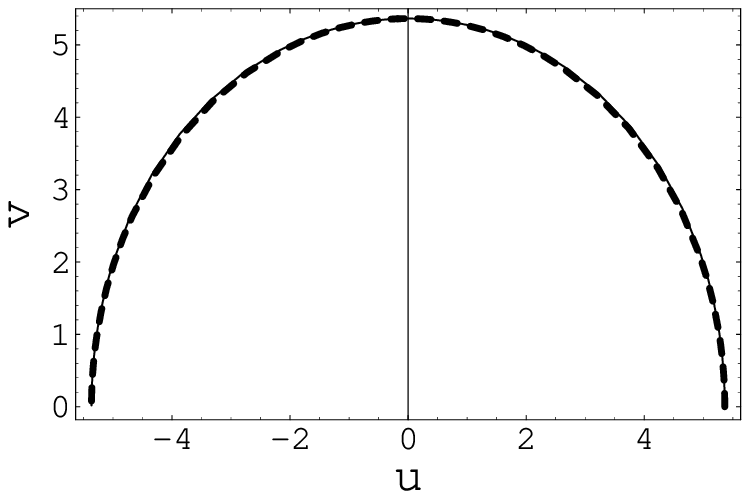}
\end{tabular}}
\caption{Left, the square of the wave packet $| \psi(u,v)|^2$ for
$A^*(n)=\frac{n\,\chi^n}{\,{\sqrt{2^n\,n!}}}e^{-\chi^2/4}$ and
$\chi=5$. Right, the classical (dashed line) and bohmian (solid
line) trajectories.} \label{fig2}
\end{figure}

Now, we can use the method that is developed in Ref.~\cite{pedram}
to construct the ``canonical" wave packets. The canonical wave
packets contain all desired properties to have a good classical and
quantum correspondence. The general wave packet which satisfies
above equation can be written as
\begin{equation}\label{psi}
\Psi(u,v)=\sum_{n=\mbox{\footnotesize{even}}} A_n
\psi_n(u)\psi_n(v)+i\sum_{n=\mbox{\footnotesize{odd}}} B_n
\psi_n(u)\psi_n(v).
\end{equation}
Since the potential term is symmetric, the eigenfunctions are
separated in two even and odd categories. The initial wave function
and its initial derivative take the form
\begin{eqnarray}
\Psi(u,0)=\sum_{n=\mbox{\footnotesize{even}}} A_n
\psi_n(u)\psi_n(0),\\
\frac{\partial\Psi(u,v)}{\partial
v}\bigg|_{v=0}=i\sum_{n=\mbox{\footnotesize{odd}}} B_n
\psi_n(u)\psi'_n(0).
\end{eqnarray}
Therefore, the $A_n$ coefficients determine the initial wave
function and $B_n$ coefficients determine the initial derivative of
the wave function. As a mathematical point of view, since the
underling differential equation (\ref{4.3}) is second order, $A_n$s
and $B_n$s are arbitrary and independent variables. On the other
hand, if we are interested to construct the wave packets which
simulate the classical behavior with known classical positions and
velocities, these coefficients will not be all independent yet. It
is obvious that the presence of the odd terms of $v$ dose not have
any effect on the form of the initial wave function but they are
responsible for the slope of the wave function at $v=0$, and vice
versa for the even terms. Near $v=0$ the differential equation
(\ref{4.3}) takes the form
\begin{eqnarray}
\left\{-\frac{\partial^2}{\partial u^2}+\frac{\partial^2} {\partial
v^2}+ {\omega}^2u^{2}\right\}\psi(u,v)=0. \label{eq10nearv0}
\end{eqnarray}
This PDE is also separable in $u$ and $v$ variables, so we can write
\begin{equation}\label{psi-separated}
\psi(u,v)=\psi(u)\chi(v).
\end{equation}
By using this definition in (\ref{eq10nearv0}), two ODEs can be
derived
\begin{eqnarray}
\frac{d^2\chi_n(v)}{d v^2}+E_n\chi_n(v)&=&0,
\label{eqseparated1}\\
\hspace{-0.6cm}-\frac{d^2\psi_n(u)}{d
u^2}+{\omega}^2u^{2}\psi_n(u)&=&E_n\psi_n(u),\label{eqseparated2}
\end{eqnarray}
where $E_n$s are separation constants. These equations are
Schr\"{o}dinger-like equations with $E_n$s as their `energy' levels.
Equation (\ref{eqseparated1}) is exactly solvable with plane wave
solutions
\begin{equation}\label{eqplanewave}
\chi_n(v)=\alpha_n\cos\left(\sqrt{E_n}\,\,v\right)+i\beta_n\sin\left(\sqrt{E_n}\,\,v\right),
\end{equation}
where $\alpha_n$ and $\beta_n$ are arbitrary complex numbers.
Equation (\ref{eqseparated2}) is Schr\"odinger equation for SHO with
the well known solutions (\ref{4.5}). Now, the general solution to
equation (\ref{eq10nearv0}) can be written as,
\begin{eqnarray}\label{psi-separated2}\nonumber
\psi(u,v)&=&\sum_{n=\mbox{\footnotesize{even}}} A^*_n
\cos(\sqrt{E_n}v)
\psi_n(u)+i\sum_{n=\mbox{\footnotesize{odd}}}B^*_n\sin(\sqrt{E_n}v)
\psi_n(u).
\end{eqnarray}
As stated before, this solution is valid only for small $v$. The
general initial conditions is
\begin{eqnarray}
\psi(u,0)=\sum_{even}A^*_n\psi_n(u)\label{eqinitial1}\\
\hspace{-0.8cm}\psi'(u,0)=i\sum_{odd}B^*_n\sqrt{E_n}\psi_n(u),\label{eqinitial2}
\end{eqnarray}
where prime denotes the derivative with respect to $v$. Obviously a
complete description of the problem would include the specification
of both these quantities. However, since we are interested to
construct the wave packet with all classical properties, we need to
assume a specific relationship between these coefficients. The
prescription is that the functional form of undetermined
coefficients i.e. $B^*_n$ for $n$ odd, are equal to the functional
form of determined coefficient i.e. $A^*_n$ for $n$ even
\cite{pedram}
\begin{equation}\label{eqcanonicalslope}
 B^*_n=A^*_n \,\,\,\,\,\, \mbox{for $n$
  odd}.
\end{equation}
Therefore, in terms of $A_n$s and $B_n$s (\ref{psi}) we have
\begin{equation}\label{eqcanonicalslope2}
 B_n=\frac{i\sqrt{E_n}}{\psi'_n(0)}\psi_n(0) A_n \,\,\,\,\,\, \mbox{for $n$
  odd}.
\end{equation}
Note that $\psi_n(0) A_n$ for $n$ odd, are defined to have the same
functional form as for $n$ even. We will see that this choice of
coefficients leads to a good classical and quantum correspondence.
Figure \ref{fig1} shows the resulting wave packet for a particular
choice of initial condition
($A^*(n)=\frac{\,\chi^n}{\,{\sqrt{2^n\,n!}}}e^{-\chi^2/4}$). These
coefficients are chosen such that the initial state consists of two
well separated peaks and this class of problems are the ones which
are also amenable to a classical description. As can be seen from
Fig.~\ref{fig1}, the wave function is smooth and its crest follows
the classical trajectory. In fact, we are free to choose any other
appropriate initial condition. Figure \ref{fig2} shows the resulting
wave packet with different initial condition. We see that this wave
packet also contain the same behavior as the previous one. Note
that, these two initial conditions correspond to two different
classical description with radii $\xi=5$ (Fig.~\ref{fig1}) and
$\xi=5.364$ (Fig.~\ref{fig2}), respectively. In next Section, to
make the connection between quantum mechanical and classical
solutions more clear, we study this issue from Bohmian point of
view.

\section{Bohmian trajectories}\label{sec4}
To make the connection between the classical and quantum results
more concrete, we can use the ontological interpretation of quantum
mechanics \cite{holland,nelson}. Moreover, since time is absent in
quantum cosmology we can recover the notion of time using this
formalism.

\begin{figure}
\centerline{\begin{tabular}{ccc}
\includegraphics[width=8cm]{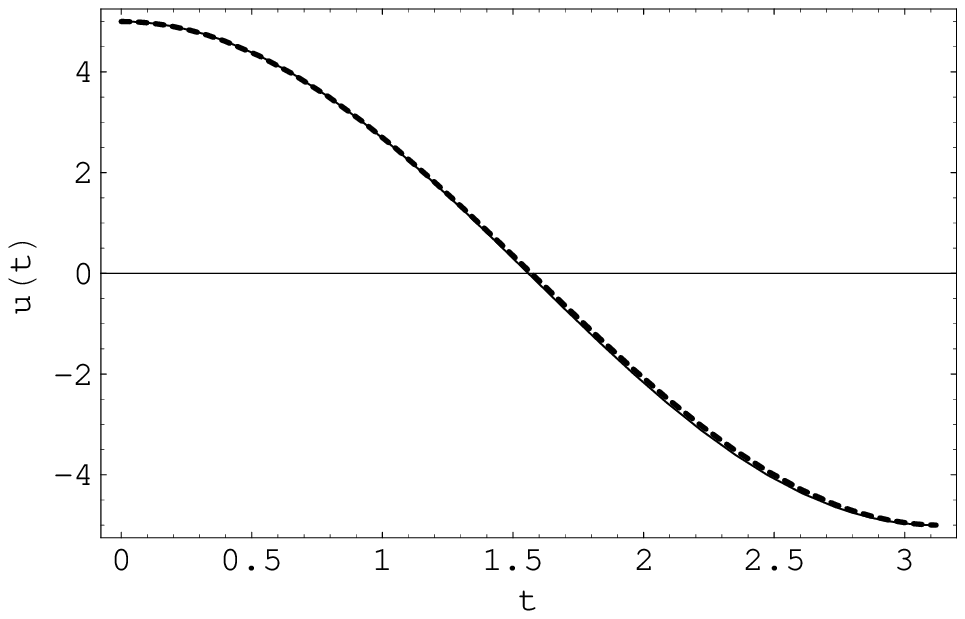}
 &\hspace{2.cm}&
\includegraphics[width=8cm]{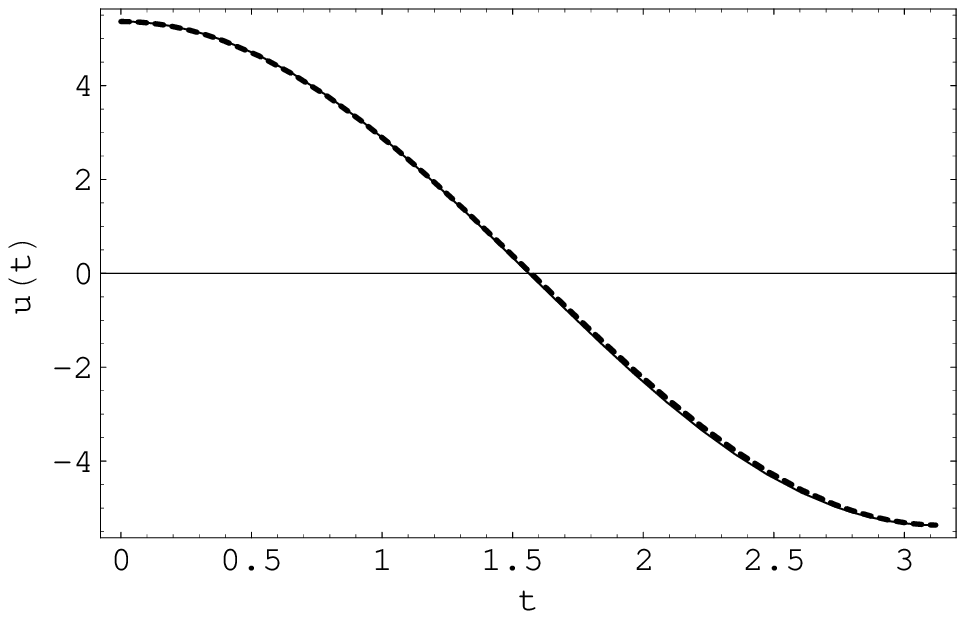}
\end{tabular}}
\caption{Plot of $u(t)$ for Classical (dashed line) and Bohmian
(solid line) trajectories where
$A^*(n)=\frac{\chi^n}{\,{\sqrt{2^n\,n!}}}e^{-\chi^2/4}$ (left),
$A^*(n)=\frac{n\,\chi^n}{\,{\sqrt{2^n\,n!}}}e^{-\chi^2/4}$ (right)
and $\chi=5$.} \label{fig2-1}
\end{figure}

\begin{figure}
\centering
\includegraphics[width=7.cm]{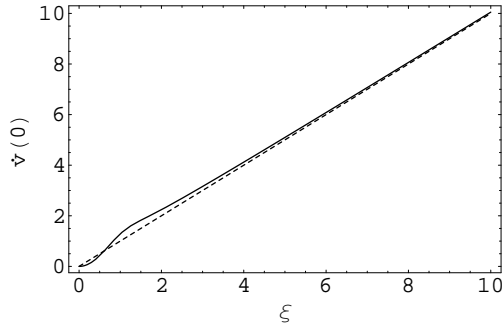}
\caption{Initial velocity ($\dot{v}(0)$) via classical (dashed line)
and de-Brogli bohm interpretation of quantum mechanics (solid
line).} \label{fig3}
\end{figure}

In ontological interpretation the wave function can be written as
\begin{equation}\label{R,S}
\Psi(u,v) = R\, e^{iS}
\end{equation}
where $R=R(u,v)$ and $S=S(u,v)$ are real functions and satisfy the
following equations
\begin{eqnarray}
\label{hje} -\frac{\partial^2 R}{\partial u^2} +\frac{\partial^2
R}{\partial v^2}+R\left(\frac{\partial S}{\partial u}\right)^2-
R\left(\frac{\partial S}{\partial v}\right)^2+ {\omega}^2(u^{2}-v^{2})R&=& 0,\\
R\frac{\partial^2 S}{\partial u^2} -R\frac{\partial^2 S}{\partial
v^2}+2 \frac{\partial R}{\partial u} \frac{\partial S}{\partial
u}-2\frac{\partial R}{\partial v} \frac{\partial S}{\partial v}&=&0.
\end{eqnarray}
To write $R$ and $S$, it is more appropriate to separate the real
and imaginary parts of the wave packet
\begin{equation}
\Psi (u,v)=x(u,v)+i y(u,v)
\end{equation}
where $x,y$ are real functions of $u$ and $v$. Using equation
(\ref{R,S}) we have
\begin{eqnarray}
R&=&\sqrt{x^2+y^2}\\
S&=&\arctan\left(\frac{y}{x}\right)
\end{eqnarray}
On the other hand, the Bohmian trajectories, which determine the
behavior of the scale factor, are governed by
\begin{eqnarray}
p_u = \frac{\partial S}{\partial u},\\
p_v = \frac{\partial S}{\partial v}.
\end{eqnarray}
where the momenta correspond to the classical related Lagrangian
($L(q)=\dot{q}^2-V(q)$). Therefore, the equations of motion take the
form
\begin{eqnarray}
\dot{u}= \frac{1}{2}\frac{1}{1+\left(\frac{y}{x}\right)^2}\frac{d}{du}\left(\frac{y}{x}\right),\\
\dot{v}=-\frac{1}{2}\frac{1}{1+\left(\frac{y}{x}\right)^2}\frac{d}{dv}\left(\frac{y}{x}\right).
\end{eqnarray}
Using the explicit form of the wave packet (\ref{R,S}), these
differential equations can be solved numerically to find the time
evolution of $u$ and $v$. In the right part of
Figs.~\ref{fig1},\ref{fig2}, we superimposed the classical and
Bohmian trajectories for two different choices of initial
conditions. The coincidence between these two trajectories is
apparent from the figure. Moreover, the obtained Bohmian position
versus time ({\it i.e.~}$u(t)$) coincide well with its classical
counterpart (Fig.~\ref{fig2-1}). In particular, Fig.~\ref{fig3}
shows the initial velocity at $v=0$ versus classical radius from
classical and de-Brogli bohm points of view. As can be seen from the
figure, the classical-quantum correspondence is manifest for large
$\xi$, where $\xi$ is the classical radius of motion. In fact, the
difference between classical and Bohmian results for small $\xi$ is
due to the interference between the parts of the wave function and
can be reduced by making the wave function more localized over the
classical path \cite{pedram}.

\section{Conclusions}\label{sec5}
We have studied $(n+1)$--dimensional cosmology with varying speed of
light. We have obtained exact solutions in both classical and
quantum levels for ($c\,$--$\Lambda$)--dominated Universe. We then
constructed the wave packets via canonical proposal which exhibit a
good classical-quantum correspondence. This method propose a
particular relation between even and odd expansion coefficients
which construct the initial wave functions and the initial
derivative of the wave functions, respectively. In other words,
canonical prescription define a particular connection between
position and momentum distributions which at the same time
correspond to their classical quantities and respect to the
uncertainty relation. We have also studied the situation using
de-Broglie Bohm interpretation of quantum mechanics. In fact,
Bohmian trajectories highly depend on the wave function of the
system and various linear combinations of eigenfunctions lead to
different Bohmian trajectories. Therefore, the inconsistency between
classical and Bohmian trajectories is natural in most cases. In this
paper, using canonical prescription, we have tried to construct the
wave packets which peak around the classical trajectories and
simulate their classical counterparts. Using Bohmian interpretation
we quantified our purpose of classical and quantum correspondence
and showed that the Bohmian positions and momenta coincide well with
their classical values upon choosing arbitrary but appropriate
initial conditions. It is worth to mention that the classical and
quantum correspondence issue has been attracted much attention in
the literature \cite{Coleman}. In particular, Hawking and Page
\cite{Page} and Kiefer \cite{Kiefer} have also been studied the same
WDW equation and discussed the situations where the resulting wave
packets exhibit classical properties. But since the Kiefer's
proposal of initial condition result in the real wave function, it
does not correspond to any classical trajectory. In summary,
canonical proposal can be considered as a general, simple and
efficient method to construct wave packets with a complete classical
behavior for various physical models where we encounter with WDW-like
equations.

\section*{Acknowledgements}
It is our pleasure to dedicate this paper to Professor Hamid Reza
Sepangi and gratefully acknowledge most useful interactions with
him.

%+Bibliography

\end{document}